\documentclass[apj]{emulateapj}

\usepackage{color}
\usepackage{graphicx,epstopdf}
\usepackage{enumerate}
\usepackage[bookmarks=true, pdfstartview={FitH},hyperfootnotes=false]{hyperref}

\newcommand{\elem}[1]{{\scriptsize~{#1}}}

\newcommand{\Chandra}{\emph{Chandra}}
\newcommand{\Suzaku}{\emph{Suzaku}}

\shorttitle{Exploring the bridge between A3556 and A3558}
\shortauthors{Ursino, E., Galeazzi, M., Gupta, A., Kelley, R. L., Mitsuishi, I., Ohashi, T., Sato, K.}

\begin{document}
\title{Exploring the bridge between A3556 and A3558 in the Shapley Supercluster}


\author{Ursino, E., Galeazzi, M.\altaffilmark{1},}
\affil{Physics Department, University of Miami, Coral Gables, FL 33155, USA}
\author{Gupta, A.,}
\affil{Department of Biological and Physical Sciences, Columbus State Community College, Columbus, OH 43215, USA} 
\affil{Department of Astronomy, The Ohio State University, 140 West 18th Avenue, Columbus, OH 43210, USA }
\author{Kelley, R. L.,}
\affil{NASA/Goddard Space Flight Center, Greenbelt, MD 20771, USA}
\author{Mitsuishi, I.,}
\affil{Division of Particle and Astrophysical Science, Nagoya University, Furo-cho, Chikusa-ku, Nagoya, Aichi 464-8602, Japan}
\author{Ohashi, T.,}
\affil{Department of Physics, Tokyo Metropolitan University, Tokyo 192-0397, Japan}
\author{Sato, K.}
\affil{Department of Physics, Tokyo University of Science, Tokyo 162-8601, Japan}
\altaffiltext{1}{corresponding author, galeazzi@physics.miami.edu}


\begin{abstract}

Looking at the region connecting two clusters is a promising way to identify and study the Warm-Hot Intergalactic Medium. Observations show that the spectrum of the bridge between A3556 and A3558 has a stronger soft X-ray emission than the nearby region. \Suzaku\ observations could not discriminate the origin of the extra emission. In this work we analyze a dedicated \Chandra\ observation of the same target to identify point sources and characterize the background emission in the bridge. We find that the count number of the point sources is much higher than average field population (using CDFS~4~Ms as a reference). Moreover, the shape of the cumulative distribution resembles that of galaxy distribution suggesting that the point sources are galaxies in a filament. The \Suzaku\ extra emission is well explained by the high abundance of point sources identified by \Chandra. Furthermore, we used optical/IR observations of point sources in the same field to estimate the density of the putative filament as $\rho\approx150 \rho_b$, below \Suzaku\ sensitivity.

\end{abstract}



\section{Introduction}
\label{introduction}
High precision cosmological surveys (\emph{Planck, ACT}, and \emph{WMAP}, see \citealt{Ade14}, \citealt{Sievers13}, and \citealt{Bennett13} for references) measured the cosmological baryonic fraction as $\Omega_b\approx0.046$. In the nearby Universe ($z<2$), instead, surveys probe only $\sim70\%$ of the baryons \citep{Fukugita98,Shull12}. Large scale hydrodynamical simulations \citep{CenOst99, Borgani04,Oppenheimer08,Schaye10} predict that a large fraction of the baryons is in the form of a filamentary gas at $10^5$~K$<T<10^7$~K and density (in terms of baryonic density) $\rho<1000\rho_b$. This gas is called the Warm-Hot Intergalactic Medium (WHIM) and it is traced by galaxies and clusters of galaxies embedded in it.

The WHIM should be visible in the UV and Soft X-ray ($\sim100-2000$~eV) bands. UV \citep{Danforth05, Danforth08, Tripp08} and Ly-$\alpha$ \citep{Richter04, Danforth10} surveys found evidence of several absorbers associated to WHIM at $T<10^6$~K. The warmer WHIM should be visible in the X-rays, but so far there are just a few \citep{Nicastro05, Fang02, Fang07, Buote09, Zappacosta10, Nicastro13} and sometimes controversial \citep{Kaastra06, Rasmussen07} detected absorbers. Recently \citet{Williams13} suggested that the absorption features in the X-ray band are not due to the WHIM but rather to intervening galaxies.

The WHIM is one of several sources that contribute to the Diffuse X-ray Background (DXB). Models predict that the WHIM contribution is $\sim10\%$ of the total DXB \citep{Phillips01, Galeazzi09}, making it difficult to observe in emission.

Filaments have been observed with \emph{XMM-Newton} between clusters A222 and A223 \citep{Werner08, Dietrich12}, in the visible (\emph{Subaru}), X-rays (\Chandra) towards the massive cluster MACS J0717.5+3745 \citep{Ebeling04, Ma09, Jauzac12}, and with combined X-ray/Infrared (\emph{ROSAT, Planck}) observations \citep{Ade13}. All these filaments, however, have temperatures of the order of 1~keV or more, too high for the WHIM. The analysis of the angular correlation function in several ``empty sky'' fields provided evidence of the WHIM emission \citep{Galeazzi09}, but no detection of individual filaments.

A region between two nearby clusters offers the best chance to detect a filament. Besides \citet{Werner08}, there are two reported \Suzaku\ observations towards the bridges between A2804 and A2811 in the Sculptor supercluster \citep{Sato10} and between A3556 and A3558 in the Shapley supercluster (\citealt{Mitsuishi12}, M12 from now on). None of them found evidence of WHIM emission but were able to set upper limits to the intensity of the O\elem{VII} and O\elem{VIII} emission lines (the best tracers of WHIM emission) and to the baryonic overdensity in the two regions.

\begin{figure}
\plotone{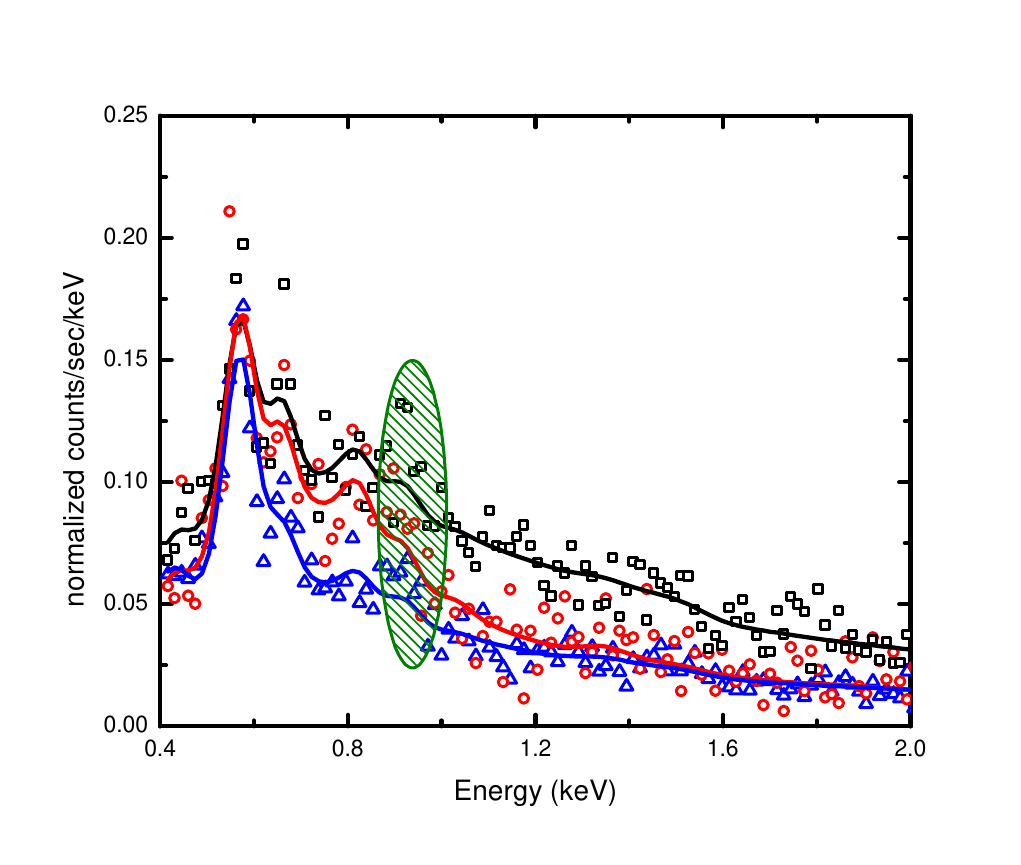}
\caption{The non-Xray background removed spectra and best fit models of the off-filament $1^\circ$ (red circles and line), off-filament $4^\circ$ (blue triangles and line), and on-filament (black squares and line) observations with \Suzaku, as reported in \citet{Mitsuishi12}. The on-filament target shows a strong excess emission that poorly fits any thermal models. The extra emission could instead be explained with excess point sources in the field. The green region highlights the Ne\elem{IX} emission line, particularly strong in the on-filament target. The model shown is the ``3T'' model from M12. (A color version of this figure is available in the online journal.) 
\label{Spectrum1}}
\end{figure}

M12 studied a bridge between clusters A3556 and A3558, in the Shapley supercluster, and two nearby targets, at $\sim1^\circ$ and $\sim4^\circ$ away from the bridge, in relatively empty regions. The $4^\circ$-offset target sets the foreground emission (due to Solar Wind Charge Exchange, Local Bubble, and Galactic Halo). There is no evidence of O\elem{VII} and O\elem{VIII} emission from the filament and an upper limit of $1.5\times10^{-7}$~photons~s$^{-1}$~cm$^{-2}$~arcmin$^{-2}$ to the O\elem{VII} intensity, corresponding to a gas at $T=3\times10^6$~K and $\rho\sim 380 (Z/0.1Z_\odot)^{-1/2} (L/3$~Mpc$)^{-1/2}$, was set. The spectrum of the on-filament target (see figure~\ref{Spectrum1}) shows extra emission and a significant enhancement at energies of Ne\elem{IX}, possibly indicating an active star forming region. The authors attempted to fit the extra emission with a $\sim2$~keV plasma, a temperature much higher than the WHIM, but without strong constraints on the fitting parameters. The 2~keV plasma model can be associated to several non-WHIM extended sources (like the outskirts of A3556 and A3558 or the plasma in the Supercluster itself). 

In our work we investigate the possibility that the extra emission comes from unresolved point sources in the on-filament target (from now on Shapley filament), if they are more abundant than in the reference offset targets and than in reported literature \citep{Bardelli96, Bardelli98}. Point sources, in fact, contribute to the spectrum with a power law with photon index $\Gamma\sim1.4$ \citep{Mushotzky00}. We explore this possibility using a dedicated 10~ks \Chandra\ \emph{ACIS-S} observation of the on-filament target of M12. The great \Chandra\ angular resolution (at the arcsecond scale) makes it possible to identify most of the point sources that contribute to the spectrum of figure~\ref{Spectrum1}. 

We describe the \Chandra\ observation and the reduction of the dataset in section~\ref{Chandra_obs}. In section~\ref{PS} we investigate the properties of the point sources distribution, their contribution to the overall spectrum of the \Suzaku\ target, and how they provide limits to the gas density. 

\section{The Chandra observation}
\label{Chandra_obs}

The \Suzaku\ observation of the Shapley filament was performed on July 10 2008 for a total of 30.2~ks of good time. The nominal position was $(\ell,b)=(311.44^\circ, 30.72^\circ)$ and the equivalent absorption density is estimated as $4.15\times10^{20}$~cm$^{-2}$. For the analysis of the \Suzaku\ observation we used the reduced dataset of M12, therefore we refer the reader to M12 for a thorough description of the data reduction. 

\begin{figure}
\plotone{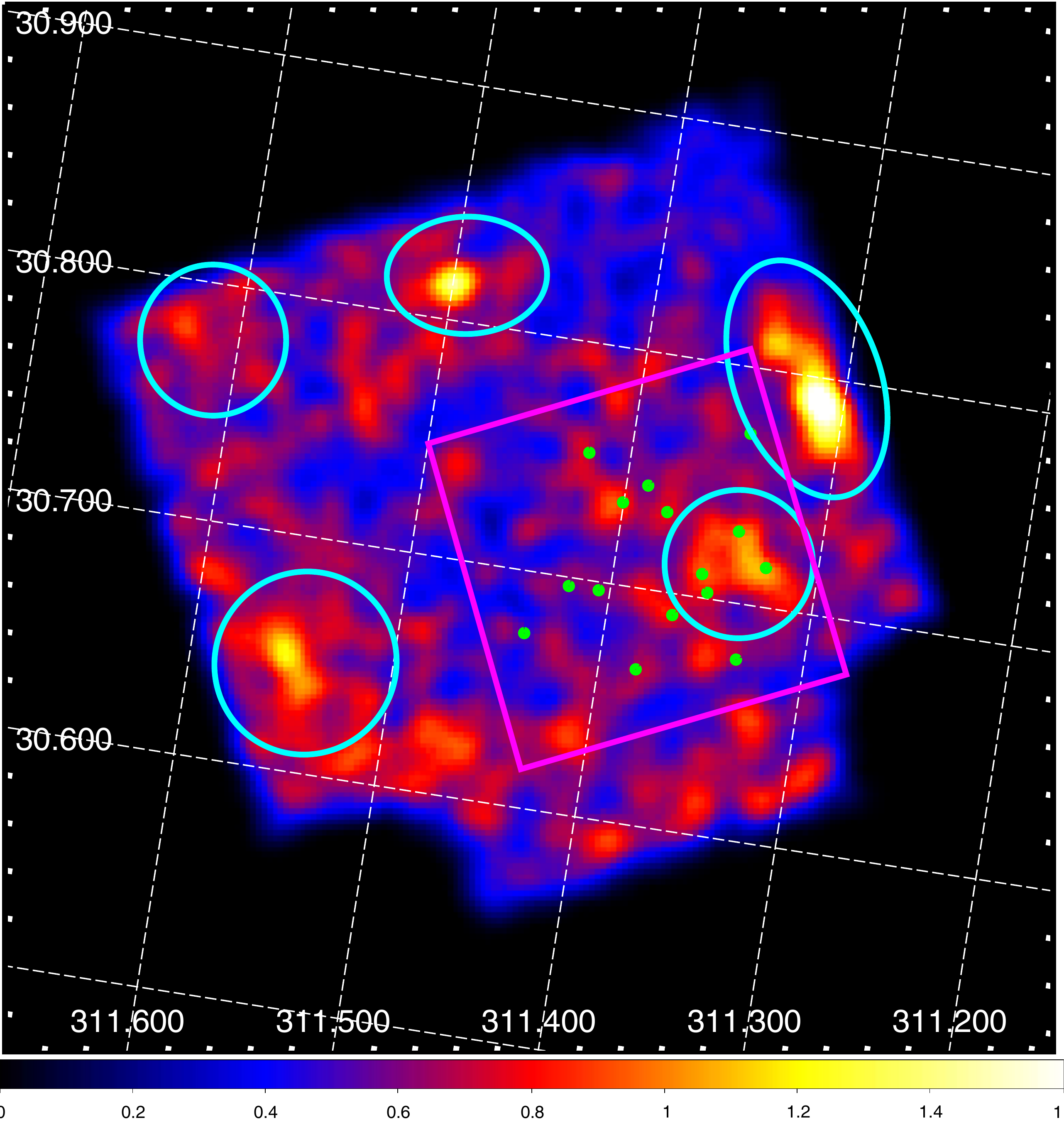}
\caption{\emph{Suzaku XIS1} image of the Shapley filament ([0.5-2.0]~keV). We highlight the \emph{Chandra} field of view (magenta square), regions removed by \citet{Mitsuishi12} (cyan ellipses), and by \emph{Chandra} (green disks). The colorbar scale is in counts~s$^{-1}$. (A color version of this figure is available in the online journal.)
\label{Map_optical}}
\end{figure}

The 10~ks \Chandra\ observation was performed on July 6 2011 towards $(\ell,b)=(311.36^\circ, 30.67^\circ)$ using the \emph{ACIS-S} setup to maximize the effective area at energies below 1~keV.  The \Chandra\ field of view covers the innermost region of the \Suzaku\ pointing, as shown in figure~\ref{Map_optical}. Within the \Chandra\ field of view there is only one of the point sources detected by \Suzaku, leaving a large area to identify extra point sources.

We filtered the dataset to remove bad time intervals plagued by flares. We generated the $[0.5-2]$~keV map for chipset 7 and the corresponding exposure map (with a reference spectrum based on the model described in \citealt{McCammon02}). We used the \emph{CIAO wavdetect} tool to identify the point sources in the map. To reduce the risk of false positives in regions with low exposure, we masked the maps in order to look for point sources only where pixels have exposure higher than $2.5\times10^6$~cm$^2$~s (equivalent to 30\% of the maximum exposure), thus reducing the solid angle of the field to $\sim0.017$~deg$^2$. The detection tool identified 15 X-ray point sources in the map, with a detection limit of $5.4\times10^{-15}$~erg~cm$^{-2}$~s$^{-1}$ and a significance threshold for the point-source detection of $10^{-6}$. 

The proper computation of the cumulative count numbers requires to estimate the completeness function. By applying the \emph{wavdetect} tool to synthetic maps of random sources at increasing SB, from $10^{-15}$ to $10^{-13}$~erg~cm$^{-2}$~s$^{-1}$, we obtained the fraction of sources detected at a given luminosity. Afterwards, we normalized the count numbers to the completeness function and to a solid angle of 1~deg$^2$.

Figure~\ref{Map_optical} shows that 4 out of the 15 \Chandra\ point sources fall in the mask set for \Suzaku\ point sources removal and one source is exactly at the boundary (we considered itas being part of the \Suzaku\ mask), therefore we extracted the stacked spectra of the 10 previously unresolved sources.

\section{The excess point sources}
\label{PS}

\subsection{Density distribution}
\label{PS_distr}

CDFS is the field with the deepest exposure in the X-rays and therefore has the best characterization of point sources density distribution. We chose CDFS~4~Ms \citep{Xue11} as a reference to test the cumulative count numbers distribution of point sources in the Shapley filament. \citet{Lehmer12} performed a very thorough analysis of the $Log N-Log S$ in the CDFS, dividing the point sources in three samples (AGN, stars, and galaxies). CDFS~4~Ms gives us the option to compare the Shapley filament with the deepest field population available and to compare the distribution of the detected point sources directly with individual populations.

We tested our method to compute the count numbers on the CDFS~4~Ms catalog and compared the results directly with \citet{Lehmer12}. Overall our simplified approach is in good agreement with the $Log N-Log S$ computed in \citet{Lehmer12}, at most our method slightly underestimates the number density by a constant close to 1 (preserving the shape). 

\begin{figure}
\plotone{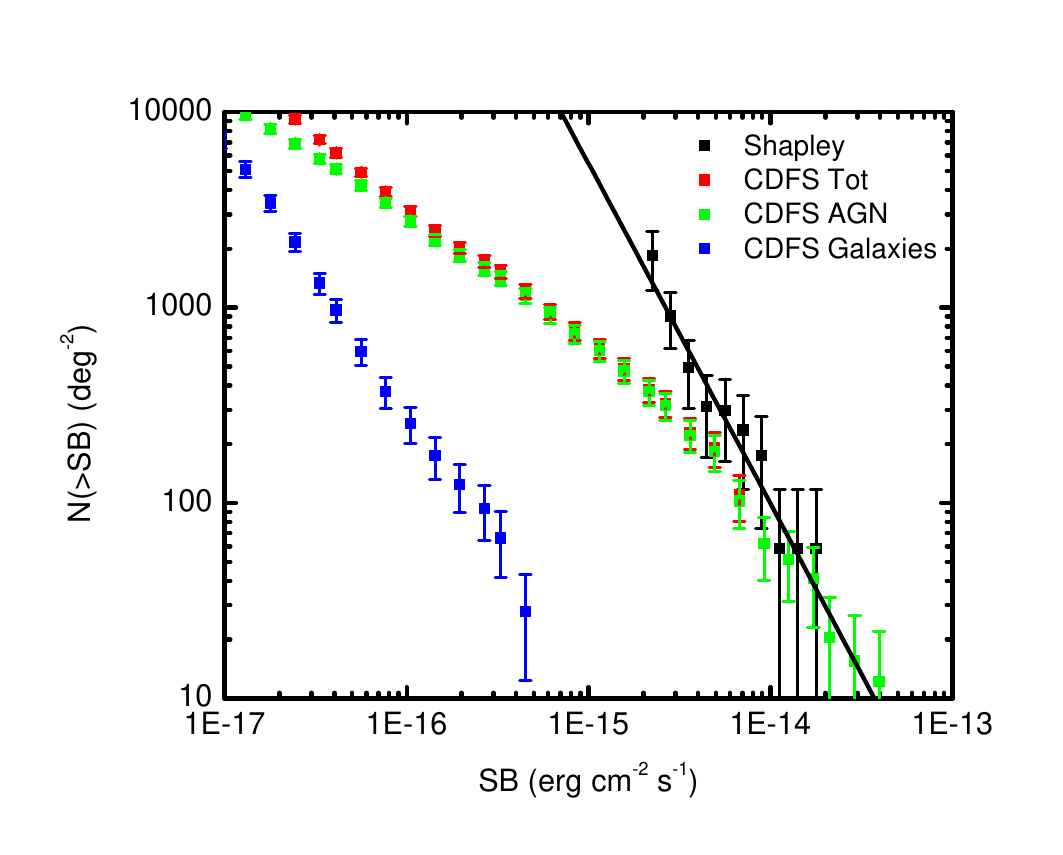}
\caption{Cumulative count numbers of sources in the Shapley filament and in CDFS. For CDFS we used \citet{Lehmer12} distributions for all point sources, for AGN  and galaxies. The solid line represents the best fit of the Shapley filament $Log N-Log S$. The total count numbers in the Shapley filament is significantly higher than the reference CDFS and has a steep slope, similar to CDFS galaxies and bright tail AGN.
\label{graph_loNloS}}
\end{figure}

In figure~\ref{graph_loNloS} we show the $Log N-Log S$ distributions of point sources in the Shapley filament and in CDFS~4~Ms, where the squares represent the data points for the Shapley dataset (black), the total population (red), the AGN population (green), and the galaxy population (blue) in CDFS, respectively. The Shapley filament clearly appears to have a much higher population density than the CDFS~4~Ms and the slope of the $Log N-Log S$ is possibly steeper than the reference populations. Using $\chi^2$ the number of point sources in the Shapley filament cannot be explained with a simple statistical fluctuation, even when cosmic variation is taken into account ($\chi^2\approx17$  with 6 d.o.f. corresponding to a likelihood of $\approx9\times10^{-3}$). Moreover, combined with the lack of excess in the ``off-filament'' targets in M12, it is a strong indication that the excess of sources come from the region between A3556 and A3558 at redshift 0.048.

Since the data are insufficient to characterize the individual point sources, we attempted to infer some information about their nature statistically, once again, by a direct comparison with the field distribution in CDFS~4~Ms. This time we looked at the individual populations identified by \citet{Lehmer12}, focusing on the galaxy and AGN populations and neglecting stars (the fraction of stars over generic point sources is very small and the chance of having stars in the Shapley field is negligible). We used the slope derived from the fit of the LogN-LogS distribution to characterize the source population. Both the Shapley distribution and the galaxy distribution were fit with a single power law. The AGN $Log N-Log S$ in \citet{Lehmer12} is well fit by a broken power law (and it is a good approximation for the distribution of all point sources, too) and we compared the Shapley population with the faint and bright tails separately. The results of the fits are shown in Table~\ref{Table_compare_chi2_Lehmer}. The comparison shows that the distribution in the Shapley pointing is incompatible with the faint tail of AGNs, but is compatible with both the distribution of galaxies and the hard tail of AGNs. However, the luminosity of point sources in the Shapley filament is at a much higher scale than that actually probed by CDFS. This can be explained if we assume that the Shapley point sources belong to the region between A3556 and A3558, i.e., they are at very small redshift ($z\approx0.048$) compared to CDFS ($z\approx1$). An excess of galaxies in the filament would, indeed, move the distribution in the LogN-LogS plot up and right (they would appear brighter). We tested this assumption by rescaling the fluxes of all CDFS galaxies to the equivalent flux at $z=0.048$. The ``rescaled'' $Log N-Log S$ is in good agreement with the Shapley sample. On the other hand, it is hard to reconcile our observation with the assumption that the excess sources are AGNs. Having already discusses how the sources must be associated with a low redshift region, the comparison with the hard tail of the AGN distribution would be inappropriate. The sources in the CDFS are at higher redshift, and a direct comparison with Shapley would require the same rescaling we tested for galaxies. The correct comparison would therefore be with the faint tail of the AGN distribution, but that is incompatible with the Shapley distribution. In conclusion, our results indicate that the excess emission detected in M12 comes from a population of galaxies belonging to the region between A3556 and A3558 and could be associated with a WHIM filament.  

The galaxy number density in the Shapley filament is slightly larger than the number density at the virial radii of A3556 and A3558 but the galaxy distribution in both clusters is elongated and the density is higher in the direction that connects the two clusters (see Fig.~1 of M12). It has to be noted that the galaxy number density at the virial radii of both A3556 and A3558 is close to 10~gal~Mpc$^{-2}$ in the NED catalog and that the number density of X-ray galaxies in the field (as shown in figures~\ref{Map_optical} and \ref{graph_loNloS}) is even larger ($\sim70$)~gal~Mpc$^{-2}$. The measured galaxy density is much larger than the average density at the cluster radius \citep{Blackburne12} where there should be at most a few gal~Mpc${-2}$ even assuming that we are at the intersection between the virial radii of two clusters. The galaxy density in the on-filament target, instead, is remarkably similar to the density reported for the filament feeding the cluster MACS J0717.5+3745 \citep{Ebeling04}, suggesting that you are probing a filamentary structure.

\begin{table}
\begin{center}
\caption{Deviation between the Shapley filament distribution and the CDFS~4~Ms \citet{Lehmer12} in terms of $\chi^2$, p-value, and $\sigma$.
\label{Table_compare_chi2_Lehmer}}
\begin{tabular}{|l|c|c|c|}
\hline
Sample & $\chi^2$ & p & $\sigma$\\
\hline
CDFS (Tot\_U$_{faint}$) & $6.90$  & $0.01136$&2.47 \\
CDFS (Tot\_U$_{bright}$) & $0.91$  & $0.341$&0.95 \\
CDFS (Galaxies) & $1.29$  & $0.255$&1.14  \\
CDFS (AGN$_{faint}$) & $7.68$  & $5.58\times10^{-3}$&2.77 \\
CDFS (AGN$_{bright}$) & $0.25$  & $0.620$&0.49 \\
\hline
\end{tabular}
\end{center}
\end{table}

\subsection{Spectral properties}
\label{PS_spectrum}
As confirmation of our results, we verified that the spectrum associated with the extra point sources is compatible with the extra emission in M12. 

\begin{figure}
\plotone{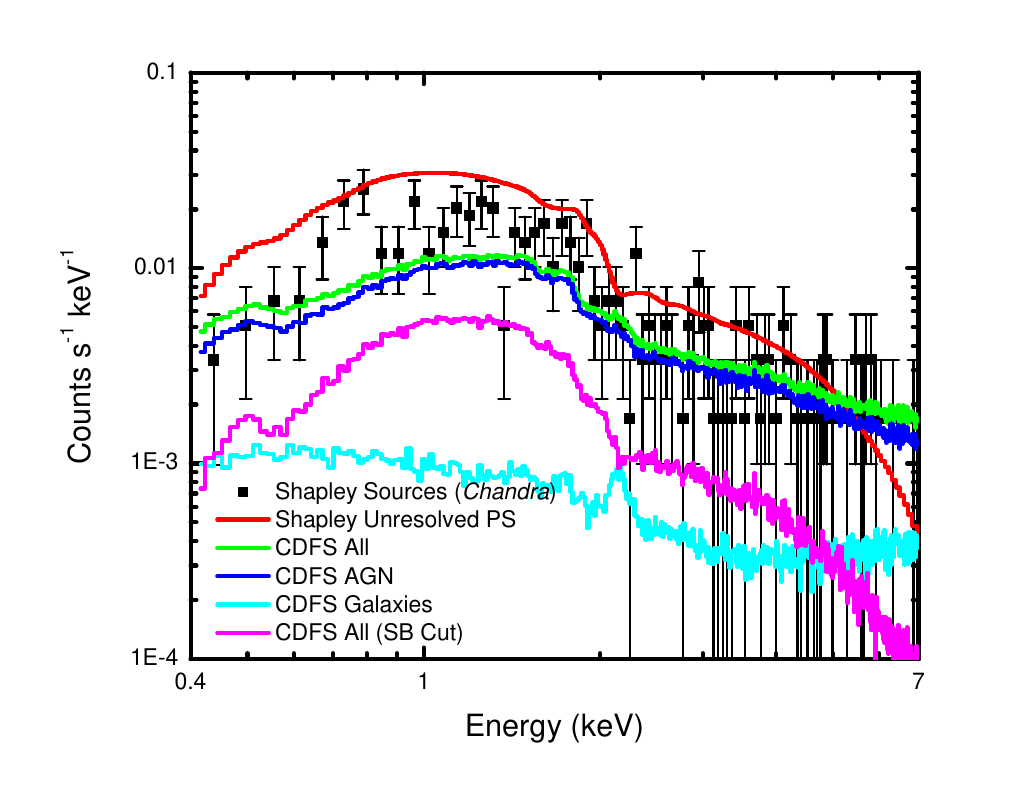}
\caption{Spectrum of the point sources resolved with \Chandra\ and of unresolved point sources in the \Suzaku\ field, compared with the spectrum of CDFS point sources. Datapoints represent the cumulative spectrum of the point sources detected in the \Chandra\ observation. The red curve represents the model spectrum of \Suzaku\ unresolved point sources, as reported in M12. The green curve represents the cumulative spectrum of point sources in the CDFS. The dark and light blue curves represent the cumulative spectra of the AGN and galaxies in the CDFS, respectively. The purple line represents the cumulative spectrum of all CDFS point sources with luminosity corresponding to the detection limit of the \Chandra\ observation of the Shapley filament. The contribution of the brightest sources in CDFS is too small to explain the extra emission in the Shapley filament. All spectra are folded through the \Chandra\ response and renormalized to a common field area.
\label{Spectrum2}}
\end{figure}

In figure~\ref{Spectrum2} we show the cumulative spectrum of the \Chandra\ point sources, the \Suzaku\ ``On-filament'' excess emission model folded through the \Chandra\ response, and the cumulative spectra extracted from the CDFS point sources. The ``On-filament'' emission (red line) is much stronger than the CDFS contribution; the emission from field unresolved sources, therefore, does not account for all the extra emission, in particular if we consider only the brightest sources (equivalent to a 10~ks \Chandra\ observation of CDFS, purple line). The cumulative spectrum of \Chandra\ point sources in the Shapley filament (black squares), instead, is in good agreement with the excess emission model. This suggests that the excess emission is likely due to a number of point sources much larger than the average field density. If the point sources in this target belonged to a field population, we would expect a spectrum in agreement with the sample made of CDFS brightest sources only. 

The fact that the measured point sources spectrum is much stronger than a field spectrum suggests that we are sampling a non-field population, possibly from a filament, leaving no room for a contribution from an extra thermal component above 1~keV.

In the analysis of the filament, M12 found evidence of excess emission near the position of the Ne\elem{IX} emission line (see figure~\ref{Spectrum1}). Fitting the emission with an extra thermal component (attributed to emission from the Supercluster) or leaving the Ne\elem{IX} abundance (\emph{vapec} with the free ``Ne'' parameter) yielded equally good parameters but there was uncertainty on the detection significance due to the uncertainties modeling the broadband excess emission. Now we can properly characterize the background, modeling it with the extra point sources we identified with \emph{Chandra}, and we have been able to estimate the significance of the Ne\elem{IX} detection at $2.4\sigma$. Even if the Supercluster origin can be ruled out for the extra emission beyond 1.2~keV (none of the control Off-targets shows sign of it in Fig.~\ref{Spectrum1}), we cannot exclude that the Supercluster is the cause of the weak Ne\elem{IX} emission.

\subsection{Limits on overdensity}
\label{overdensity}

In their work, M12 used the \Suzaku\ X-ray observation to set upper limits to the gas density toward the line of sight (overdensity$<400$). In order to estimate the gas density in the targeted region, we adopted a different approach and used data from other wavelengths. Because of the poor correspondence between X-ray point sources and the NED catalog, we selected all the $0.046<z<0.050$ NED galaxies nearby (15') our field of view to estimate the mass of baryons in galaxies, and from that we extrapolated the total baryonic mass. From the catalog, we obtained a total of 29 galaxies with photometry in the J and K bands and, in same cases, also in the B and V bands (plus 6 galaxies with no photometry).  

We estimated $M_\star$ in galaxies using the measured magnitudes and following \citet{Longhetti09} for the galaxies in the K and V bands and \citet{McGaugh14} for the B and V bands. The total star mass in the field is $M_\star\sim1.2\times10^{12} M_\odot$. By adopting the closed-box atomic gas-only model of \citet{Baldry08} and after a 15\% correction to account for unresolved galaxies \citep{Baldry08} and the 6 galaxies with no photometry, we estimated that the baryonic mass due to galaxies is $M_\star\sim1.4\times10^{12} M_\odot$.

In order to estimate the volume sampled, we assumed that the NED galaxies belong to a filament. We studied a section of the filament extracting a cylinder of depth 3.0~Mpc (\citealt{Hoshino10}, M12) and radius $\sim0.87$~h$^{-1}_{73}$~Mpc (15'), with depth directed along the line of sight. For the cylinder depth, we chose to use the size of the diameter of A3556 (the smaller of the two clusters). The total volume of the cylinder is $V\approx7.1$~Mpc$^3\approx2.1\times10^{68}$~m$^3$ (for the remainder of this section all computations already include $h$ dependency). This sets the density at $\rho_{gal}\approx2.8\times10^{-26}$~kg~m$^{-3}$, equivalent to a baryonic overdenstiy of $\sim70$. As a comparison, the density measured using weak lensing analysis of the filament feeding MACS J0717.5+3745 is $\rho=(206\pm46) \rho_b$ \citep{Jauzac12}.

To verify that the on-filament target indeed has more galaxies than regions away from clusters and filaments, we applied the same approach to the two control fields of M12 (Offset-1 and Offset-4). Offset-1 has only 1 galaxy in the field of view, but no photometry. We used a larger NED field of view (20' of radius), therefore including three more galaxies, almost equally spaced from the center, in order to obtain upper limits to density. This way we computed an overdensity limit of  $4 \rho_b$. In the field of view of Offset-4, instead, there are no galaxies so that we had to extend the NED field of view up to 30' of radius to find two galaxies and set an upper limit of $2 \rho_b$. 

According to estimates of the baryon budget \citep{Fukugita98, Fukugita04, Shull12}, galaxies make for $\sim7\%$ of the baryons at low redshift ($z<1$), the CGM for $~5\%$, and ICM for $\sim4\%$. Cold gas is $\sim1.7\%$, photoabsorbed (Ly$\alpha$) is $\sim28\%$, WHIM (OVI and Ly$\alpha$) for $\sim30\%$, and missing baryons are $\sim29\%$. In these works, the galaxy contribution corresponds to the stellar mass fraction and the neutral gas well within the galactic structure. The CGM, instead, accounts for the gas that extends for several hundreds of parsecs around the galaxies and, within the large uncertainties of its estimate, has the same order of magnitude of the stellar mass contribution. In our investigation, we compute the halo mass as a fraction ($\sim15\%$) of $M_\star$, much less than the ratio between CGM and stellar mass, indicating that the gas fraction thus estimated corresponds to the gas that more actively takes part into star formation. In our attempt to estimate the density in the region targeted by the \Suzaku\ observation, therefore, we cannot consider only the stellar mass fraction plus the halo mass, we also need to include a separate component that accounts for the gas in the CGM and/or in the WHIM closer to galaxies and clusters (since at present time there are very large uncertainties over the boundaries between these to phases). Assuming that the mass in the CGM is the same as the mass in galaxies and that there can be extra contribution from the WHIM, and considering that the part of the galaxies are indeed located well within the virial radii of A3556 and A3558, it is reasonable to set an upper limit to the density of the filament as $\rho\approx150 \rho_b$.

\acknowledgments

This work has been supported in part by SAO award \#GO1-12179X. The authors would like to thank R. K. Smith and W. Liu for the useful discussion and suggestions.

\end{document}